\documentclass[twocolumn,showpacs,preprintnumbers,amsmath,amssymb]{revtex4}
\usepackage{graphicx}
\usepackage{dcolumn}
\usepackage{bm}
\usepackage{hyperref}

\newcommand{\bef}{\begin{figure}}
\newcommand{\eef}{\end{figure}}

\newcommand{\be}{\begin{equation}}
\newcommand{\ee}{\end{equation}}
\newcommand{\bea}{\begin{eqnarray}}
\newcommand{\eea}{\end{eqnarray}}
\begin{document}


\title{Study of bulk properties of the system formed in U+U collisions at $\sqrt{s_{\mathrm NN}}$ =~2.12~GeV using JAM model}

\author{Aswini Kumar Sahoo$^{1}$, Xionghong He$^{2}$, Yasushi Nara$^{3}$ and Subhash Singha$^{2}$}
\affiliation{$^{1}$Department of Physics, Indian Institute of Science Education and Research, Berhampur, India\\
$^{2}$ Institute of Modern Physics Chinese Academy of Sciences, Lanhzou 730000, China\\
$^{3}$ Akita International University, Yuwa, Akita-city 010-1292, Japan}

\begin{abstract}
The Lanzhou Cooling-Storage-Ring facility is set to conduct experiments involving Uranium-Uranium collisions at the center of mass energies ranging from 2.12 to 2.4 GeV. Our investigation is focused on various bulk observables, which include charged particle multiplicity ($N_{\text{ch}}$), average transverse momentum ($\langle p_{\text{T}}\rangle$), initial eccentricity ($\epsilon_{n}$), and flow harmonics ($v_{n}$), for different orientations of U+U collisions within the range of $0^{\circ} < \theta < 120 ^{\circ}$ at $\sqrt{s_{\mathrm NN}} = 2.12$ GeV ($p_{\mathrm lab}$ = 500 MeV). Among the various collision configurations at this energy, the tip-tip scenario emerged with the highest average charged particle multiplicity, denoted as $\langle N_{\text{ch}} \rangle$. Notably, both the second and third-order eccentricities, $\epsilon_{2,3}$, revealed intricate patterns as they varied with impact parameter across distinct configurations. The tip-tip configuration displayed the most pronounced magnitude of rapidity-odd directed flow ($v_{1}$), whereas the body-body configuration exhibited the least pronounced magnitude. Concerning elliptic flow ($v_{2}$) near mid-rapidity ($\eta < 1.0$), a negative sign is observed for all configurations except for the side-side exhibited a distinctly positive sign. Within the spectrum of configurations, the body-body scenario displayed the highest magnitude of $v_{2}$. For reaction plane correlated triangular flow ($v_{3}$), the body-body configuration emerged with the largest magnitude while the side-side exhibited the smallest magnitude. Our study seeks to establish a fundamental understanding of various U+U collision configurations in preparation for the forthcoming CEE experiment. 
 
\end{abstract}
\pacs{25.75.Ld}
\maketitle

\section{INTRODUCTION}
Experiments involving heavy-ion collisions are designed to delve into the behavior of strongly interacting matter comprising quarks and gluons, a realm governed by Quantum Chromodynamics (QCD)~\cite{Adams:2005dq,Adcox:2004mh,BRAHMS:2004adc,Harris:1996zx,Muller:2012zq,Braun-Munzinger:2007edi}. Under conditions characterized by low temperature and low density, quarks and gluons are confined within hadrons. A primary objective of heavy-ion collision experiments is the exploration of the QCD phase diagram, which is depicted in the temperature ($T$) and baryon chemical potential ($\mu_{B}$) plane. In particular, the Beam Energy Scan (BES) program at the Relativistic Heavy Ion Collider (RHIC) facility is executed by colliding gold nuclei with a center-of-mass energy of $\sqrt{s_{\mathrm NN}}$ = 3.0–-27 GeV, corresponding to $\mu_{B}$ values spanning from about 720--155 MeV. The overarching goal is to investigate the QCD phase diagram and discern potential indicators of a QCD phase transition. Concurrently, the Cooling-Storage-Ring (CSR) facility, situated within the Lanzhou Institute of Modern Physics, is poised to conduct an External Target Experiment called CEE~\cite{Hu:2019mgr,Zhu:2021soc,Wang:2022evq,Liu:2023xhc,Zhang:2023hht,Guo:2023yal}. This experiment entails Uranium-Uranium collisions at $\sqrt{s_{\mathrm NN}}$ = 2.12–2.4 GeV and is designed to explore the QCD phase diagram under conditions characterized by higher baryon densities and lower temperatures. 

The uranium nucleus is of special interest due to its prolate-type deformed shape. Furthermore, different angles between the two colliding U nuclei can lead to special orientations during collisions, such as tip-tip, body-body, and side-side discussed in~\cite{Haque:2011aa}. The difference in initial configurations can lead to different patterns in bulk observables, such as collective flow coefficients. In fact, it is worth noting that U+U collisions have been conducted by STAR experiment at RHIC at $\sqrt{s_{\mathrm NN}}$ = 193 GeV~\cite{STAR:2015mki,STAR:2021twy,STAR:2022nvh}. There have been many attempts to differentiate between initial configurations using experimental measurements and various theoretical models~\cite{Haque:2011aa,Haque:2019vgi}. Furthermore, the deformation of the U nucleus presents a very unique chance for investigating nuclear structures, such as quadrupole ($\beta_{2}$) or octupole ($\beta_{4}$) deformations, through relativistic heavy-ion collisions~\cite{Giacalone:2021udy,Jia:2021wbq,Jia:2021qyu,Jia:2021wbq,Jia:2022qgl,Nie:2022gbg,Zhang:2022fou,Magdy:2022cvt}. This has been demonstrated at both RHIC and LHC energies~\cite{ATLAS:2022dov,Jia:IS2021talk}. Indeed, the impact of nuclear deformation has been theoretically examined in heavy-ion collisions occurring at intermediate energy ranges, typically within the range of a few GeV. Particular emphasis has been placed on comprehending the influence of collision geometry, nuclear symmetry energy, and the presence of a deformed neutron skin~\cite{Xu:2012ys,Liu:2023qeq,Fan:2023vly}. 

The primary focus of this paper is to study the bulk observables, namely charged particle multiplicity ($N_{\text{ch}}$), average transverse momentum ($\langle p_{\text{T}} \rangle$), initial eccentricity ($\epsilon_{n}$), and flow harmonics ($v_{n}$), for different orientations of U+U collisions at $\sqrt{s_{\mathrm NN}}$ = 2.12 GeV. The selections of specific orientations are discussed in the next section.  

\begin{figure}
\begin{center}
\includegraphics[scale=0.4]{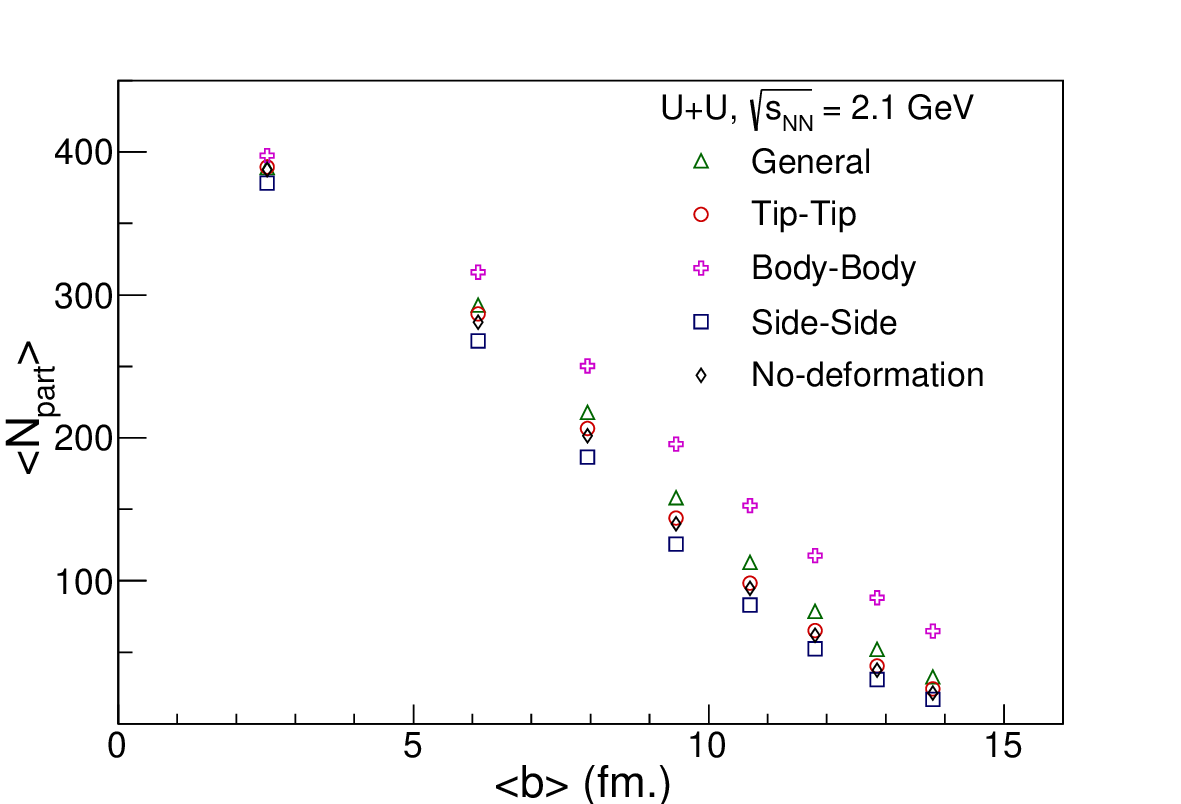}
\caption{Average of participating nucleons ($\langle N_{\mathrm part} \rangle$) as a function of impact parameter in U+U collisions at $\sqrt{s_{\mathrm NN}}$ = 2.12 GeV for different orientations of colliding nuclei from JAM model.}
\label{fig1_npart}
\end{center}
\end{figure} 

\section{The JAM Model}
In this study, we employ the Jet AA Microscopic Transport Model (JAM), a specialized hadronic transport model designed for the simulation of relativistic nuclear collisions~\cite{Nara:1999dz,Hirano:2012yy,Hirano:2012kj,Isse:2005nk,Nara:2019qfd,Nara:2020ztb,Nara:2021fuu}. The model initiates the simulation by randomly determining the initial positions of each nucleon, drawing from the distribution of nuclear density. Inside this model, the nucleon density distribution of the Uranium nucleus is parameterized by a deformed Woods-Saxon profile~\cite{Hagino:2006fj},
\begin{equation}
\rho = \frac{\rho_{0}}{1+\exp[(r-R^{\prime})/a]},\\
\end{equation}
\begin{equation}
R^{\prime} = R_{0}\big[ 1 + \beta_{2}Y_{2}^{0}(\theta) + \beta_{4}Y_{4}^{0}(\theta)   \big],
\end{equation}

where $\rho_{0}$ represents the standard nuclear density, $R^{\prime}$ signifies the nuclear radius, $a$ stands for the surface diffuseness parameter, and $Y_{l}^{m}(\theta)$ denotes spherical harmonics. In our analysis, we have adopted specific values: $R_{0}$ = 6.802 fm and $a$ = 0.54 fm. The parameters $\beta_{2}$ and $\beta_{4}$ correspond to the quadrupole and octupole deformations of U nuclei, respectively, with values set at 0.28 and 0.093~\cite{Haque:2019vgi}. Subsequent nuclear collisions are represented as the cumulative outcome of independent binary interactions between hadrons. Consequently, JAM encompasses the entire process, from the initial phase to the final state interactions within the hadronic gas phase. JAM offers two distinct modes: the cascade mode~\cite{Nara:1999dz,Hirano:2012yy,Hirano:2012kj} and the mean-field mode~\cite{Isse:2005nk,Nara:2019qfd,Nara:2020ztb,Nara:2021fuu}. In the cascade mode, each individual hadron is advanced in a manner analogous to its behavior in a vacuum, moving freely until it encounters other hadrons and experiences collisions. On the other hand, the mean-field mode incorporates nuclear equation-of-state effects through a momentum-dependent potential that influences the propagation of particles. Utilizing the mean-field approach, calculations have effectively explained the flow measurements from low-energy Au+Au collisions at RHIC, while the cascade mode proved inadequate in reproducing the experimental results. We employ JAM in its mean-field mode~\cite{Nara:2020ztb}: the relativistic quantum molecular dynamics based on the relativistic mean-field (RQMD.RMF), where we use the MD2 equation of state (EoS)~\cite{Nara:2020ztb}. The MD2 EoS has the nuclear incompressibility parameter $K = 380$ MeV, which successfully explains flow measurements at lower RHIC energies\cite{STAR:2021ozh}. In the JAM simulation, we have set the evolution time to 50 fm/$c$, using a time step of 1 fm/$c$. We have checked that results remain consistent when employing JAM up to 100 fm/$c$, except in cases of peripheral collisions. In the peripheral bins, we observed changes in results, likely stemming from prolonged interactions among particles in spectator-dominated regions ($|\eta| > 1.0$) compared to the participant-dominated regions near mid-rapidity ($|\eta| < 0.5$). This is an interesting feature that can emerge in heavy ion collisions at an intermediate energy range, where the interaction between spectators and participants plays a significant role~\cite{Sood:2006mly,Kumar:2012yu}. This aspect will be investigated in future work.

To select different angular orientations of the Uranium nucleus, we vary the polar angle ($\theta_{\mathrm t,p}$) and the azimuthal angle ($\phi_{\mathrm t,p}$) relative to the nucleus's symmetry axis for both the projectile and target nuclei. Here, the subscripts $t$ and $p$ stand for the projectile and target, respectively. These orientations are categorized as tip-tip, body-body, and side-side, as detailed in table~\ref{tab1_orientation}. The configuration categorized as general allows all possible orientations of $\theta_{\mathrm t,p}$ and $\phi_{\mathrm t,p}$, which is similar to conditions in real experiments.

\begin{table}
\caption{The polar angle ($\theta_{\mathrm t,p}$) and azimuth angle ($\phi_{\mathrm t,p}$) relative to the nucleus's symmetry axis for both the projectile and target nuclei for different orientation of uranium nuclei used in this JAM simulation.}
\begin{tabular}{|c|c|c|c|c|}
\hline
Orientations &  $\theta_\mathrm{p}$ &  $\phi_\mathrm{p}$&  $\theta_\mathrm{t}$& $\phi_\mathrm{t}$\\ 
\hline
 tip-tip  & 0 & 0--$2\pi$ & 0 & 0--$2\pi$\\
\hline
body-body   & $\pi$/2 & 0 & $\pi$/2 & 0\\
\hline
side-side   & $\pi$/2 & $\pi$/2 & $\pi$/2 & $\pi$/2\\
\hline
general   & 0--$\pi$ & 0--$2\pi$ & 0--$\pi$ & 0--$2\pi$\\
\hline
\end{tabular}
\label{tab1_orientation}
\end{table}

\begin{table}
\caption{Average of participating nucleons ($\langle N_{\mathrm part} \rangle$) in U+U collisions at $\sqrt{s_{\mathrm NN}}$ = 2.12 GeV for different orientations of colliding nuclei from JAM model.}
\begin{tabular}{|c|c|c|c|c|}
\hline
$\langle b \rangle$ (fm/c)&  (general)&  (tip-tip)&  (body-body)& (side-side)\\
\hline
 2.52  & 389.07 & 389.70 & 397.57 & 378.37\\
 6.1   & 292.90 & 286.92 & 315.75 & 267.81 \\
 7.95  & 217.90 & 206.52 & 250.18 & 186.65 \\
 9.45  & 158.04 & 143.81 & 195.80 & 125.81 \\
 10.7  & 113.00 & 98.26 & 152.58 & 82.86 \\
 11.8  & 78.68 & 65.07 & 117.7 & 52.56 \\
 12.85  & 51.84 & 40.53 & 88.20 & 30.74 \\
 13.8  & 32.85 & 24.35 & 64.86 & 17.16\\
\hline
\end{tabular}
\label{tab2_npart}
\end{table}

\section{Results and Discussion}
We perform JAM simulations comprising approximately 1 million events for each of the previously mentioned configurations, covering an impact parameter range from 0.0 to 15.0 fm. Figure~\ref{fig1_npart} illustrates the variation in the average number of participating nucleons ($\langle N_{\mathrm part} \rangle$) as a function of impact parameter. The $\langle N_{\mathrm part} \rangle$ values are also tabulated in~table~\ref{tab2_npart}. Within the array of configurations studied, it is noteworthy that the body-body scenario exhibits the highest $\langle N_{\mathrm part} \rangle$ values, while conversely, the side-side configuration demonstrates the lowest values. The tip-tip and general configurations fall between these two extremes. 

\begin{figure}
\begin{center}
\includegraphics[scale=0.4]{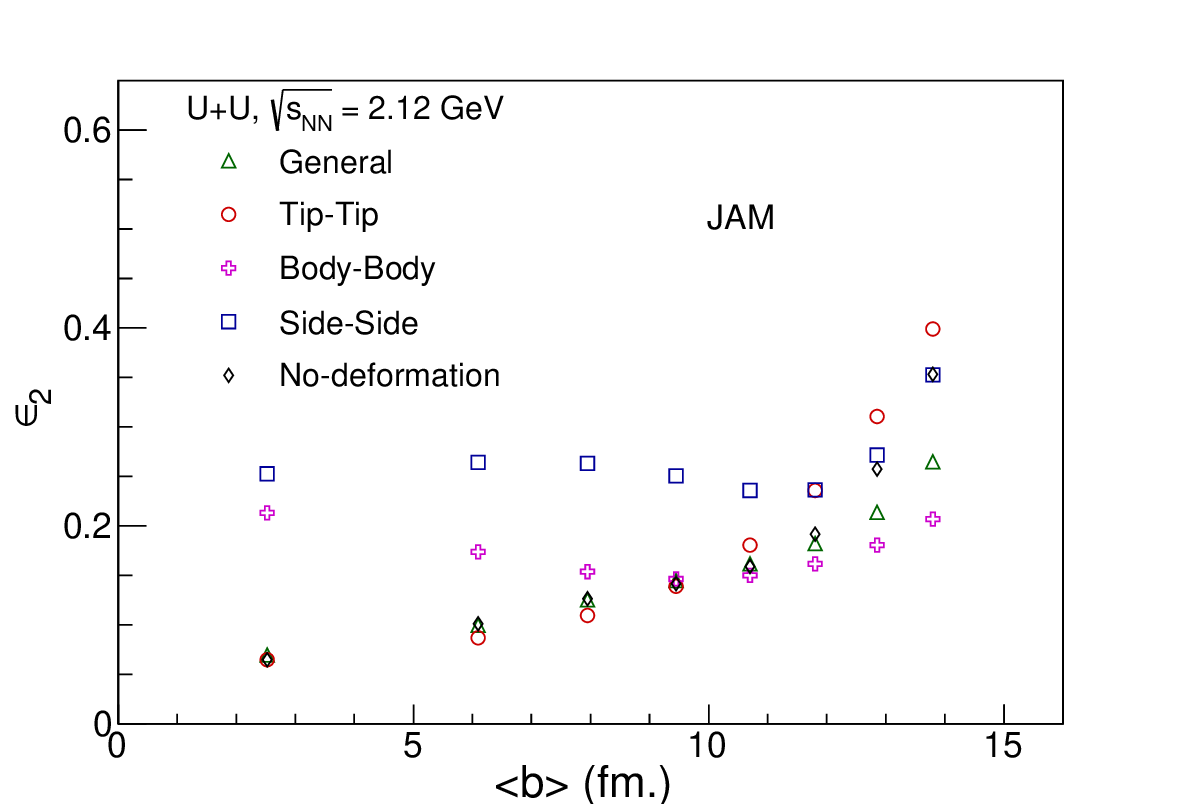}    
\includegraphics[scale=0.4]{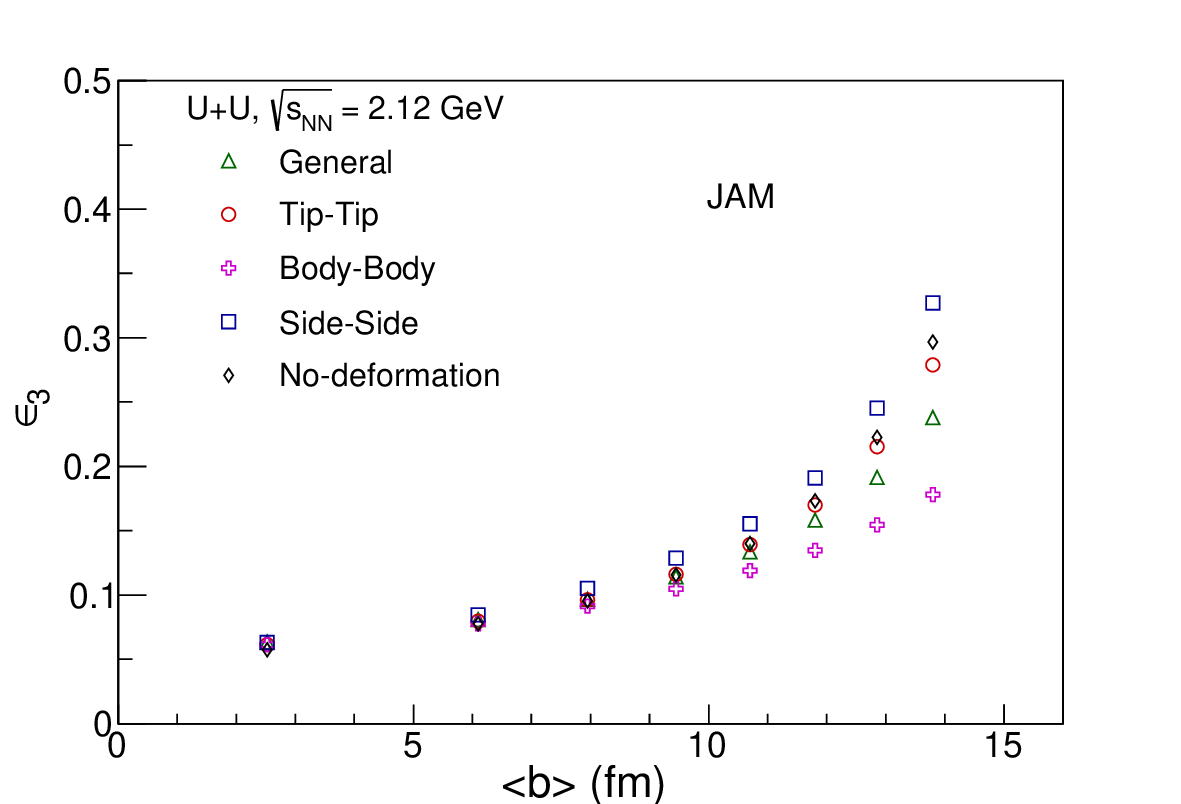}
\caption{Eccentricity of participant nucleons ($\epsilon_{2}$) as a function of impact parameter in U+U collisions at $\sqrt{s_{\mathrm NN}}$ = 2.12 GeV for different orientations of colliding nuclei using JAM model.}
\label{fig2_e23cen}
\end{center}
\end{figure}

Within the JAM framework, the initial geometric eccentricities of various configurations can be calculated from the participating nucleons following~\cite{Alver:2010gr}:
\begin{equation}
\epsilon_{n} = \frac{\sqrt{ \langle  r^{2}\cos (n \phi_{\mathrm part}) \rangle^{2} +  \langle  r^{2}\sin (n \phi_{\mathrm part}) \rangle^{2}}}{\langle r^{2} \rangle},
\end{equation}
where $r$ and $\phi_{\mathrm part}$ are the polar coordinates of participating nucleons in JAM. The angular brackets, $\langle \; \rangle$, signify that we compute averages within each event. The top and bottom panels in Fig.~\ref{fig2_e23cen} depict the variations of $\epsilon_{2}$ and $\epsilon_{3}$, respectively, stemming from different orientations as functions of impact parameter. Regarding $\epsilon_{2}$, an intricate pattern emerges. Specifically, the side-side configuration exhibits the most substantial magnitude. The body-body configuration showcases a non-monotonic trend, while the tip-tip and unconstrained configurations follow a similar trajectory. In the case of $\epsilon_{3}$, all orientations manifest a consistent increase from central to peripheral collisions, with a distinct ordering pattern pronounced in mid-central and peripheral collisions: \\side-side $>$ tip-tip $>$ general $>$ body-body.

\begin{figure}
\begin{center}
\includegraphics[scale=0.4]{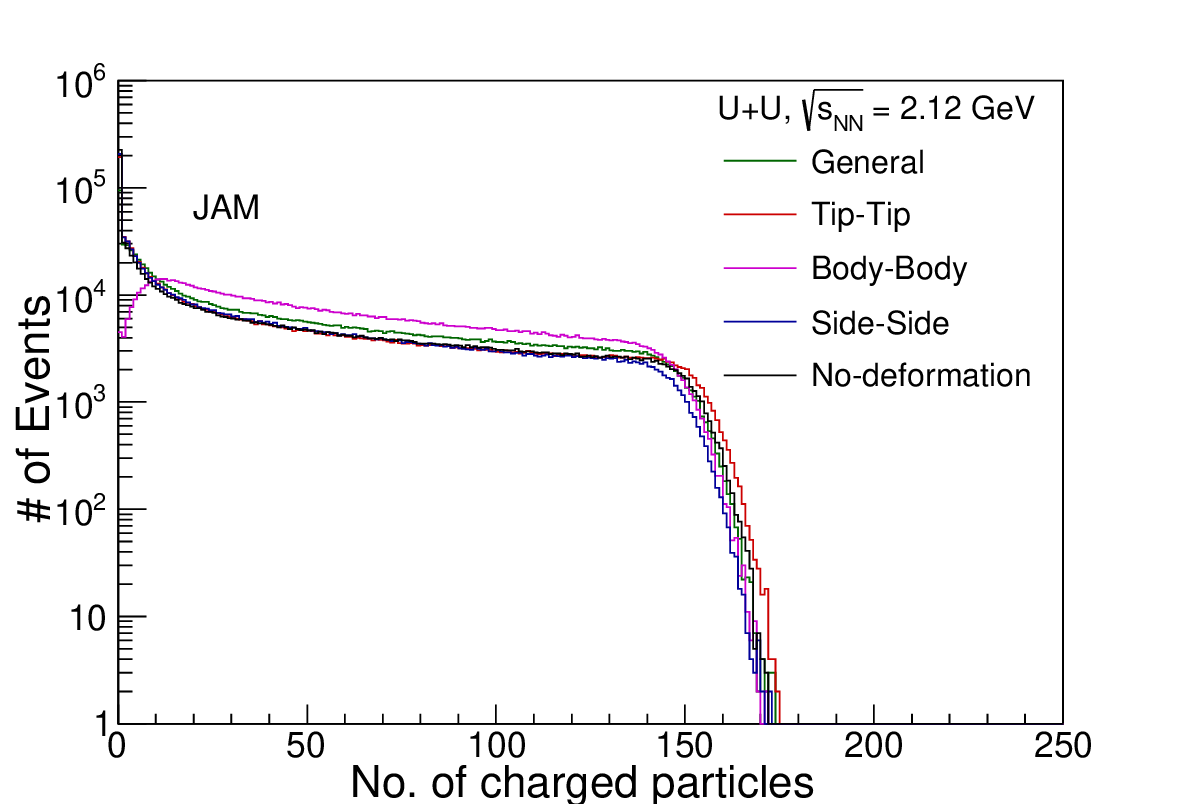}
\includegraphics[scale=0.4]{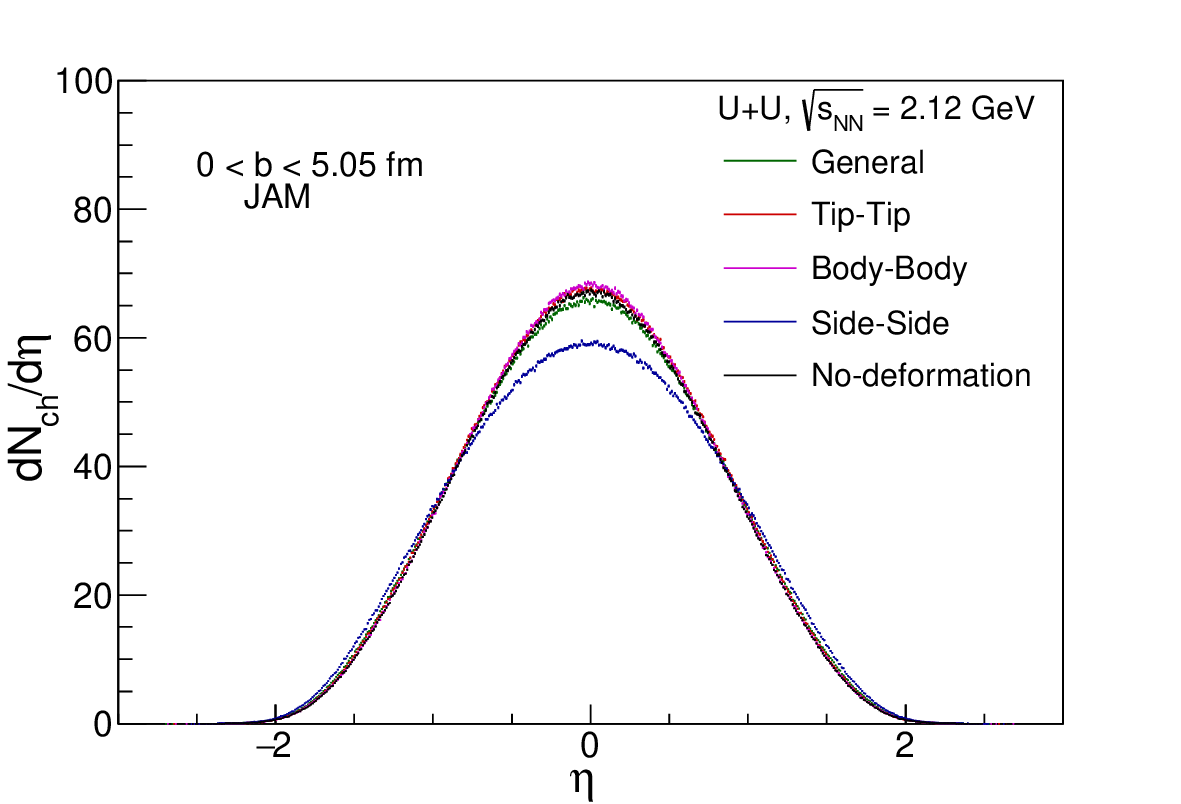}
\caption{(Top) Probability distribution of the total number of charged particles ($N_{ch}$) in U+U collisions at $\sqrt{s_{\mathrm NN}}$ = 2.12 GeV for different orientations of colliding nuclei from JAM model. (Bottom) Pseudo-rapidity density distribution of charged particles ($dN_{ch}/d \eta$) in U+U collisions at $\sqrt{s_{\mathrm NN}}$ = 2.12 GeV for different orientations of colliding nuclei using JAM model.}
\label{fig3_Nch_dNdeta}
\end{center}
\end{figure}

The top panel in Fig.~\ref{fig3_Nch_dNdeta} shows the charged particle multiplicity distributions ($N_{ch}$) from different orientations of U+U collisions. Largely, the pattern is the same for all configurations except body-body. The maximum value of $N_{ch}$ is attained in tip-tip configuration. The bottom panel in Fig.~\ref{fig3_Nch_dNdeta} presents the pseudo-rapidity density ($dN_{ch}/d \eta$) distribution of charged particles in central ($0.0 < b < 5.05$ fm) U+U collisions for different orientations. Among different configurations, there is no appreciable change in the shape of $dN_{ch}/d \eta$. 

\begin{figure}
\begin{center}
\includegraphics[scale=0.4]{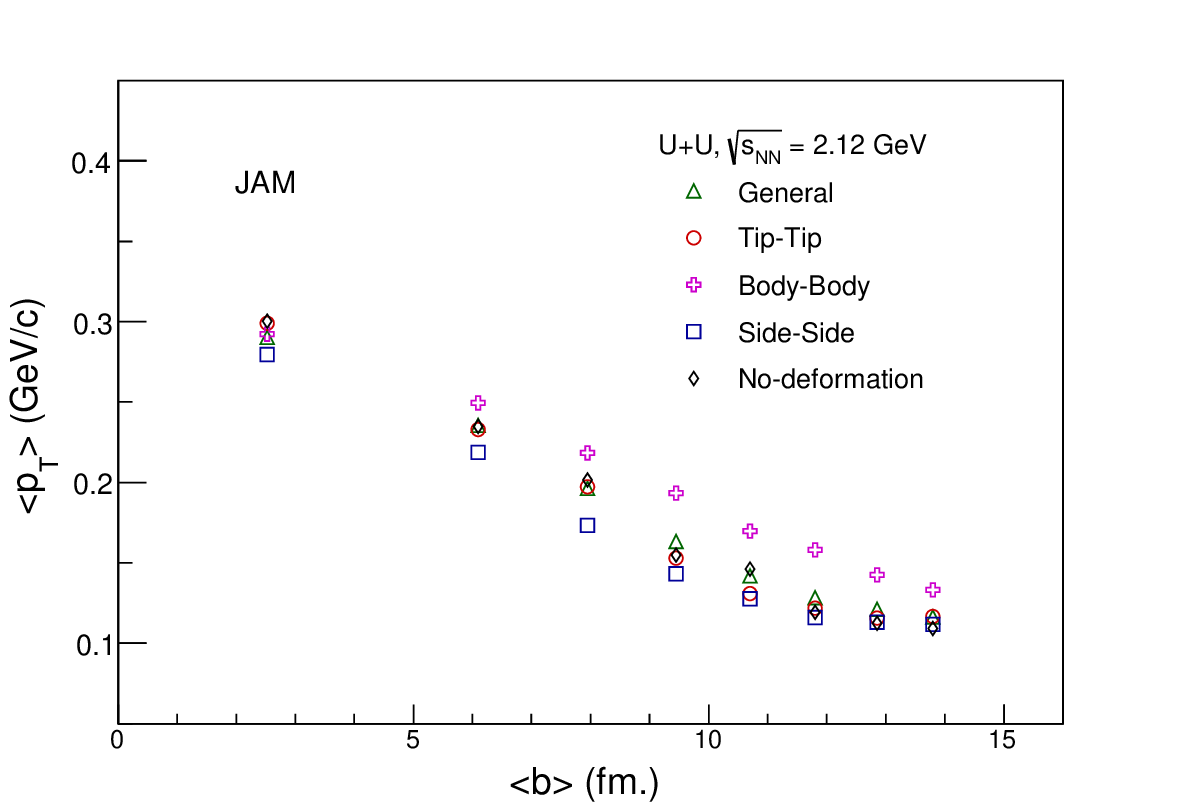}
\includegraphics[scale=0.4]{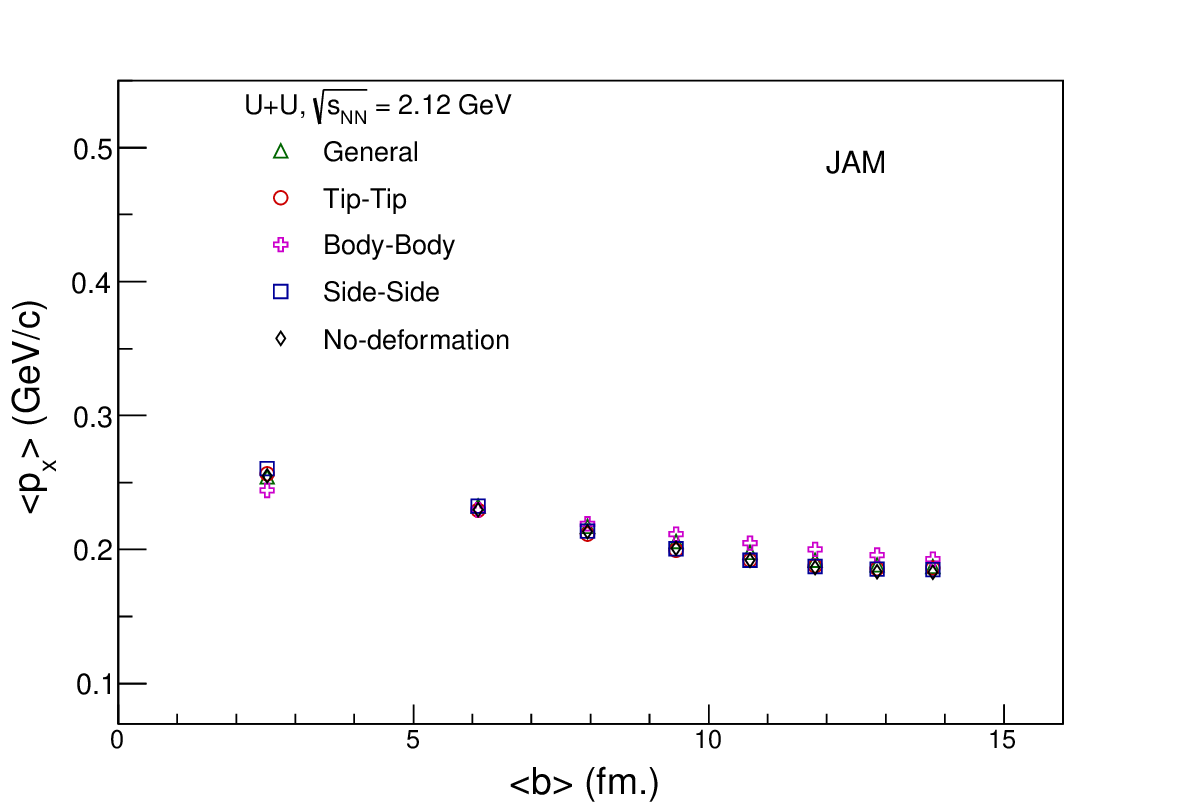}
\includegraphics[scale=0.4]{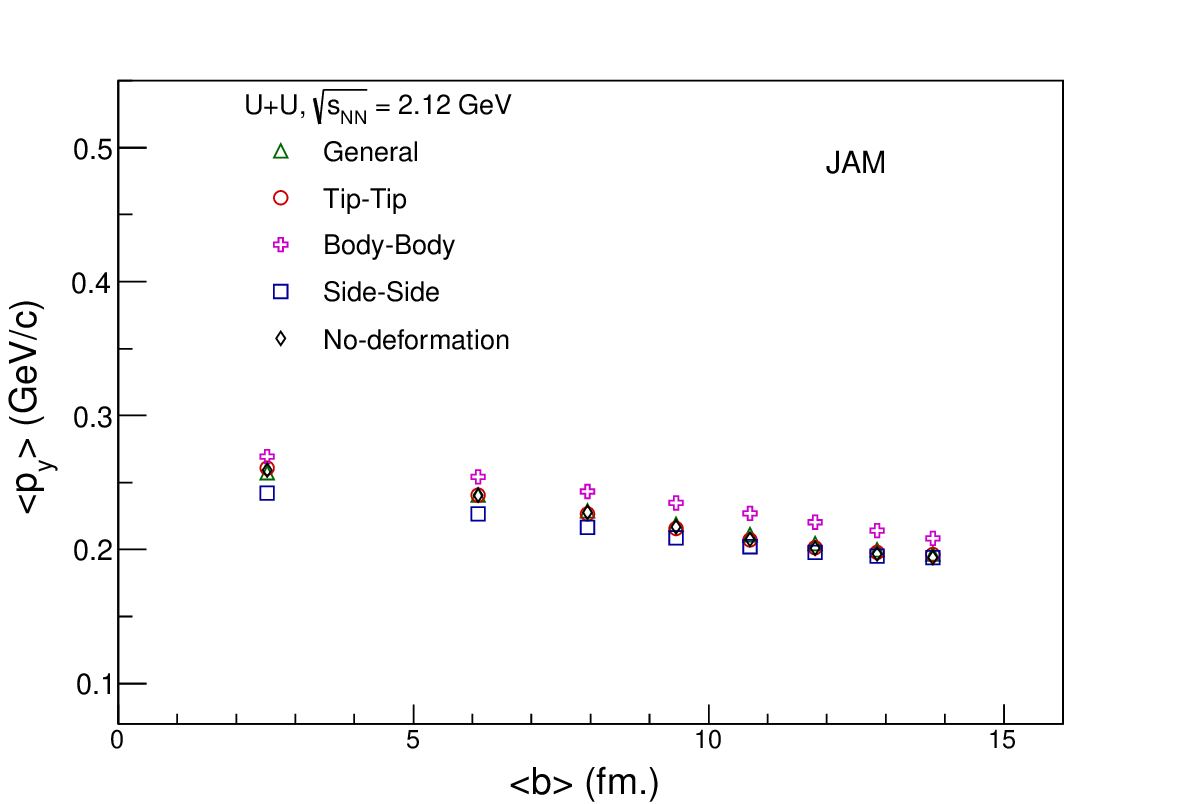}
\caption{Average momentum of charged particles ($\langle p_{\mathrm T} \rangle$, $\langle p_{\mathrm x} \rangle$ and $\langle p_{\mathrm y} \rangle$) in U+U collisions at $\sqrt{s_{\mathrm NN}}$ = 2.12 GeV for different orientations of colliding nuclei using JAM model.}
\label{fig4_mpt}
\end{center}
\end{figure}

The top panel in Fig.~\ref{fig4_mpt} displays the averaged transverse momentum of all charged particles as a function of the impact parameter. The centrality-related trends appear to be consistent across all configurations, with one notable exception: in mid-central collisions, the body-body configuration exhibits an enhanced $\langle p_{\mathrm T} \rangle$ in comparison to the rest of the configurations. This observation is quite unexpected. In fact, the  $\langle p_{\mathrm T} \rangle$ is found to be higher in the tip-tip configuration compared to other orientations of U+U nuclei at $\sqrt{s_{\mathrm NN}}$=200 GeV~\cite{Haque:2011aa}, and this can be attributed to the smaller transverse size and larger binary collisions in this configuration. In order to comprehend our observation, we studied  $\langle p_{\mathrm x} \rangle$ and $\langle p_{\mathrm y} \rangle$ for various orientations and presented in middle and bottom panels in Fig.~\ref{fig4_mpt}. Our findings reveal that $\langle p_{\mathrm y} \rangle$ is consistently larger in body-body configurations compared to other setups. This observation is also substantiated by another intriguing result, which shows that the magnitude of $v_{2}(\eta)$ in the body-body configuration (shown in Fig.~\ref{fig6_v2etapt}) is the most negative, likely as a consequence of enhanced nuclear shadowing.

\begin{figure}
\begin{center}
\includegraphics[scale=0.4]{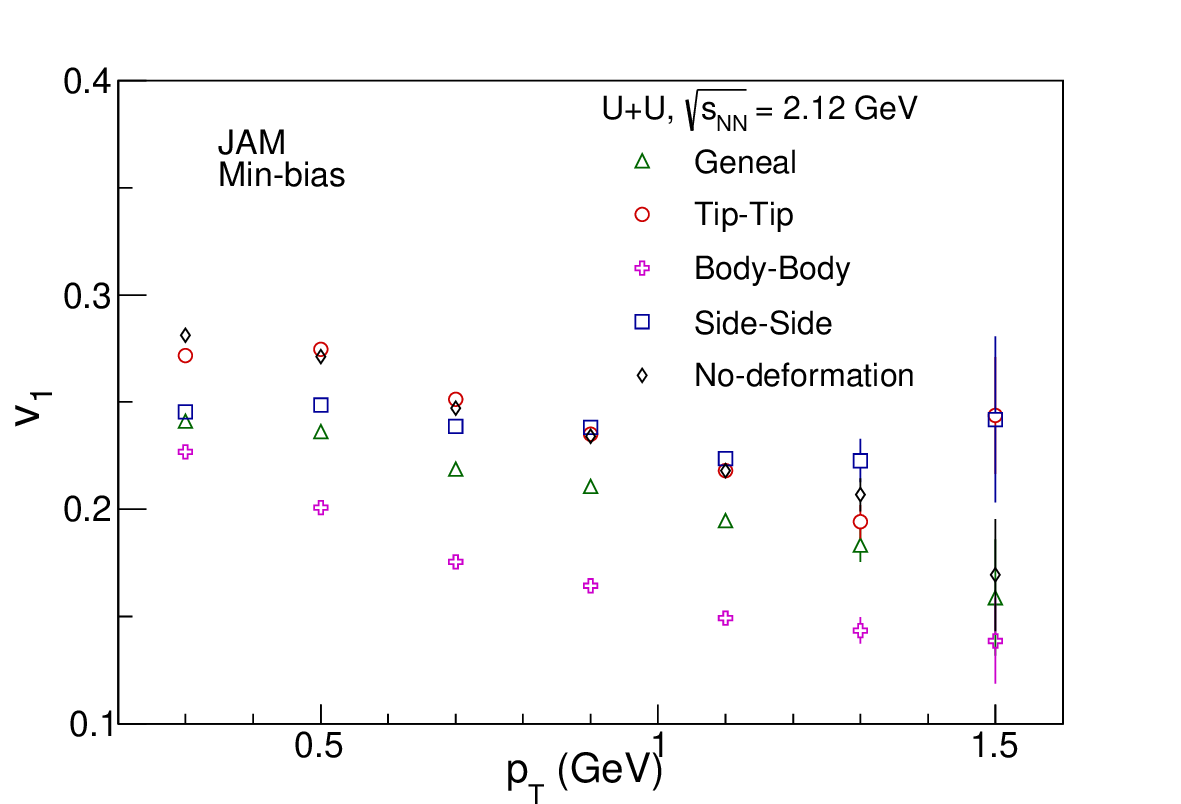}
\includegraphics[scale=0.4]{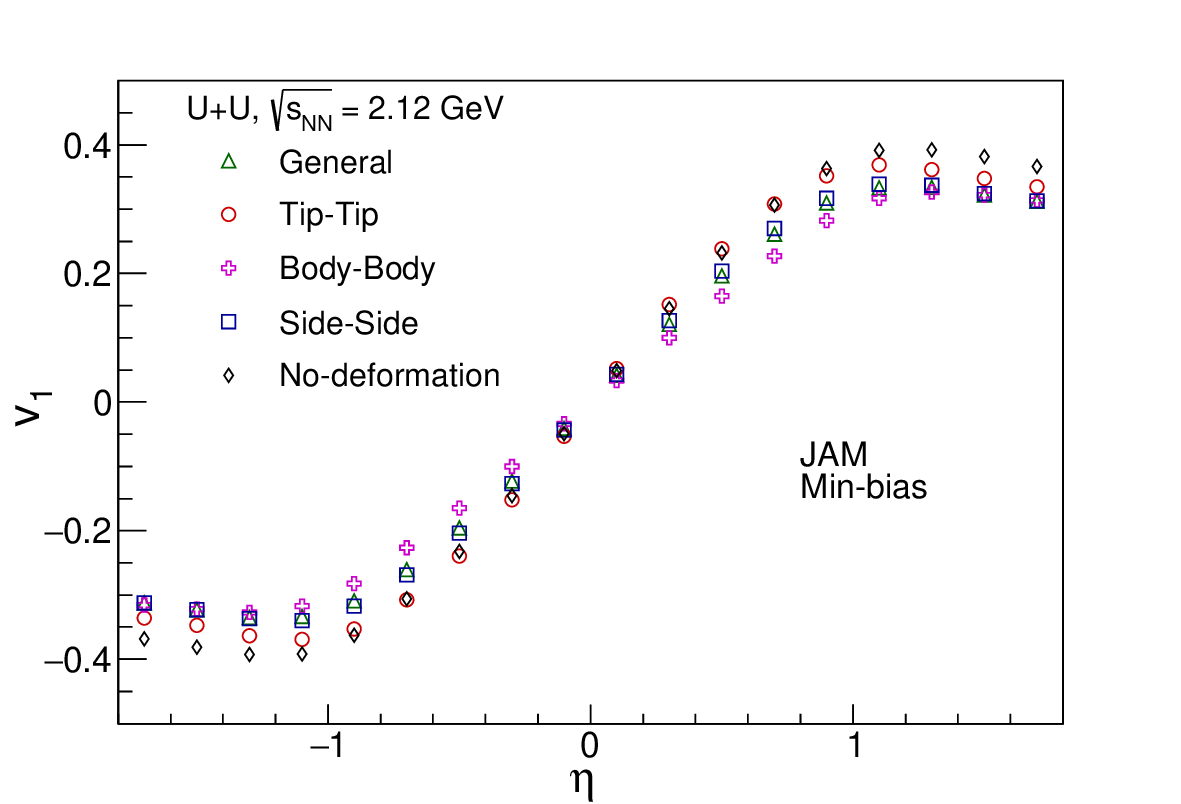}
\caption{Directed flow ($v_{1}$) of charged particles as a function of $p_{\mathrm T}$ (top panel) and $\eta$ (bottom panel) in minimum bias U+U collisions at $\sqrt{s_{\mathrm NN}}$ = 2.12 GeV for different orientations of colliding nuclei using JAM model.}
\label{fig5_v1etapt}
\end{center}
\end{figure}

\begin{figure}
\begin{center}
\includegraphics[scale=0.4]{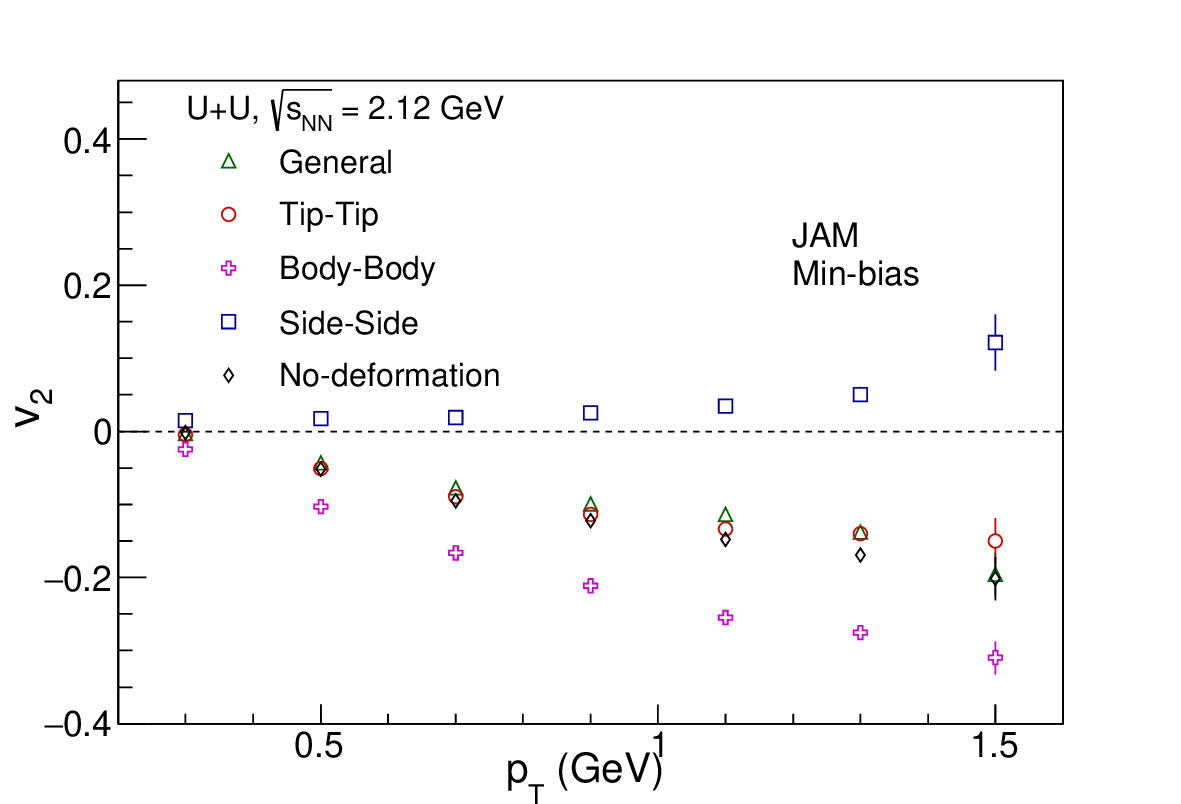}
\includegraphics[scale=0.4]{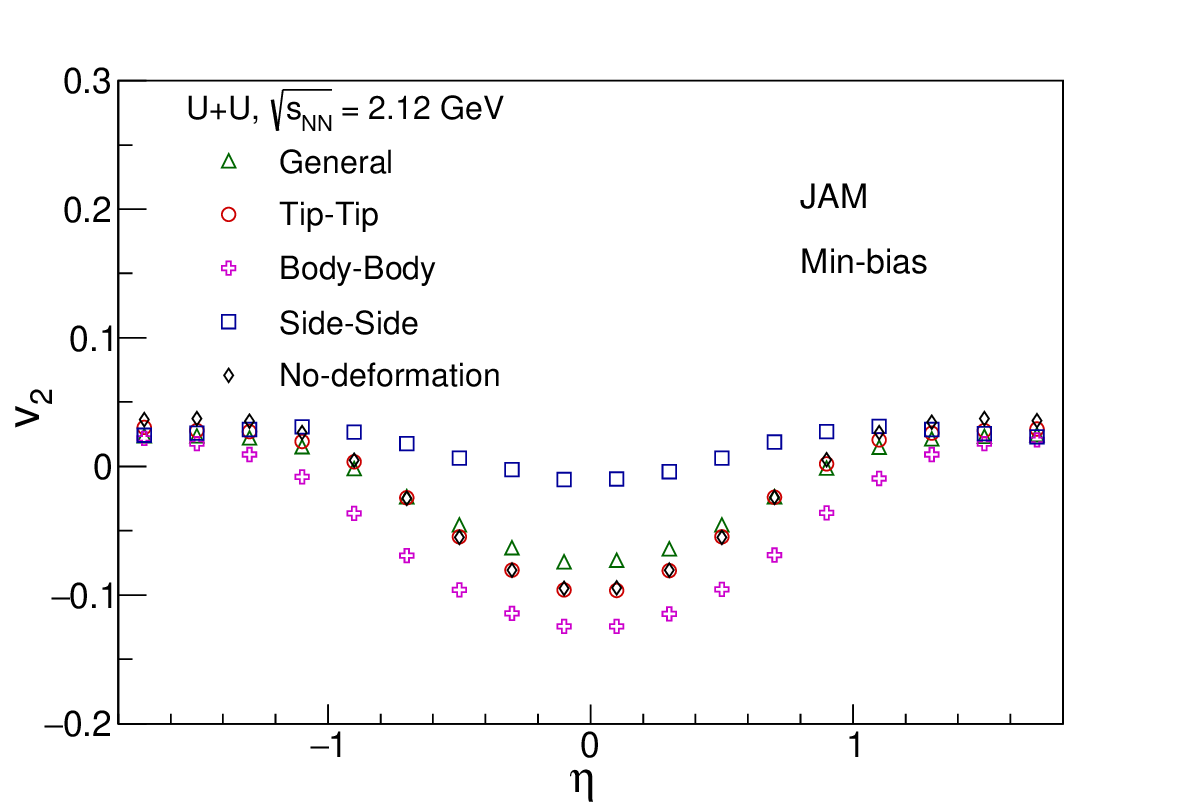}
\caption{Same as Fig.~\ref{fig5_v1etapt} but for elliptic flow ($v_{2}$).}
\label{fig6_v2etapt}
\end{center}
\end{figure}

Next we calculate the flow harmonics ($v_{n}$) which is defined by~\cite{Ollitrault:1992bk,Poskanzer:1998yz}
\begin{equation}
v_{n} = \langle \langle \mathrm cos \; n(\phi-\Psi_{\mathrm RP}) \rangle \rangle, 
\end{equation}
where $\phi$ is the azimuthal angle of the produced particles and $\Psi_{\mathrm RP}$ is the reaction plane. The $v_{n}$ is averaged over all events and all particles. In JAM simulation, the reaction plane is along the impact parameter direction, hence $\Psi_{\mathrm RP}=0$. The top and bottom panel of Fig.~\ref{fig5_v1etapt} shows the rapidity-odd directed flow ($v_{1}$) of charged particles as a function of $p_{\mathrm T}$ and $\eta$
for minimum bias U + U collisions at $\sqrt{s_{NN}}=2.12\,\mathrm{GeV}$. The tip-tip configuration displays the largest $v_{1}$-slope, while the body-body configuration displays the smallest slope.

The top and bottom panel of Fig.~\ref{fig6_v2etapt} presents the elliptic flow ($v_{2}$) of charged particles as a function of $p_{\mathrm T}$ and $\eta$. The elliptic flow ($v_{2}$) displays a notably intricate pattern that varies across different U+U configurations. Typically, within this range of beam energy, the sign of $v_{2}$ is negative, primarily due to the dominance of in-plane flow over out-of-plane flow due to the shadowing by the spectators. 
Near mid-pseudorapidity, such negative $v_{2}$ trend is particularly pronounced in the general configurations, including tip-tip and side-side. However, there is a distinct departure from this pattern in the case of the side-side configuration, where there is a marked shift with a positive sign for $v_{2}$.
This is a very striking observation; in the future event-shape engineering technique will be employed to explore the possibility of disentangling side-side configurations from an unbiased case.

\begin{figure}
\begin{center}
\includegraphics[scale=0.4]{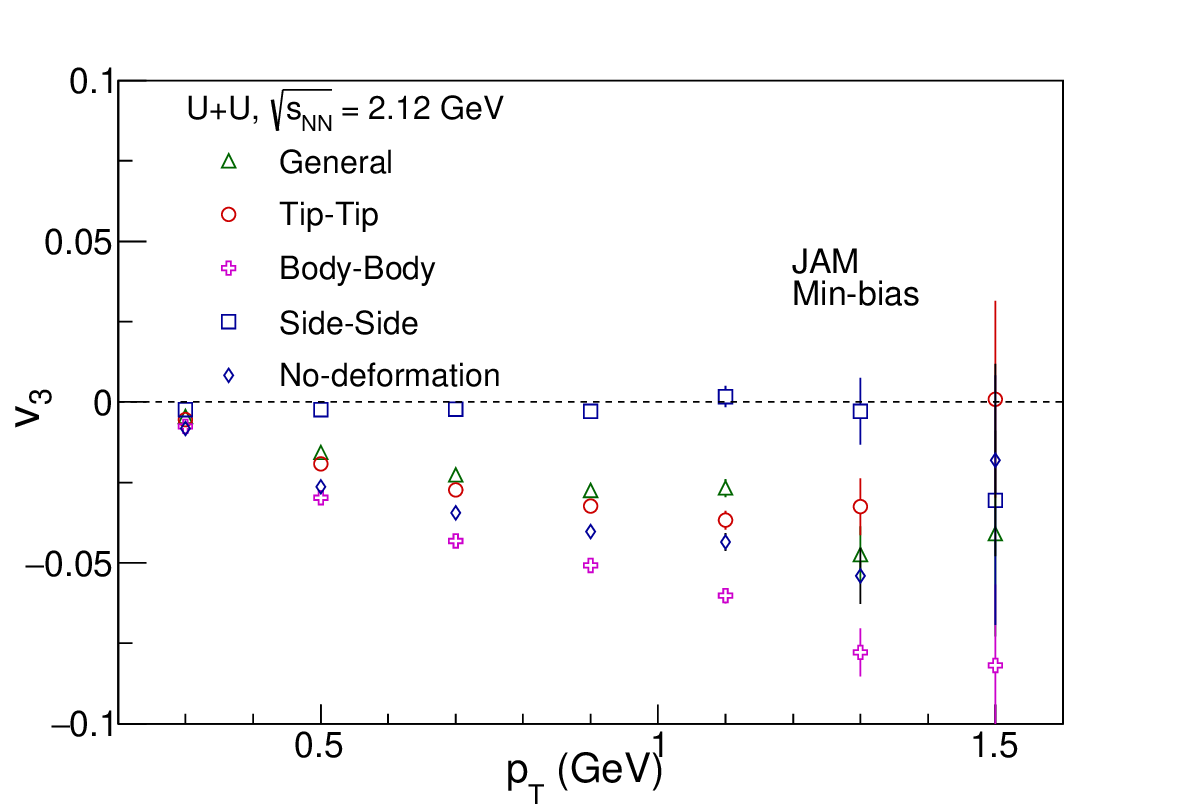}
\includegraphics[scale=0.4]{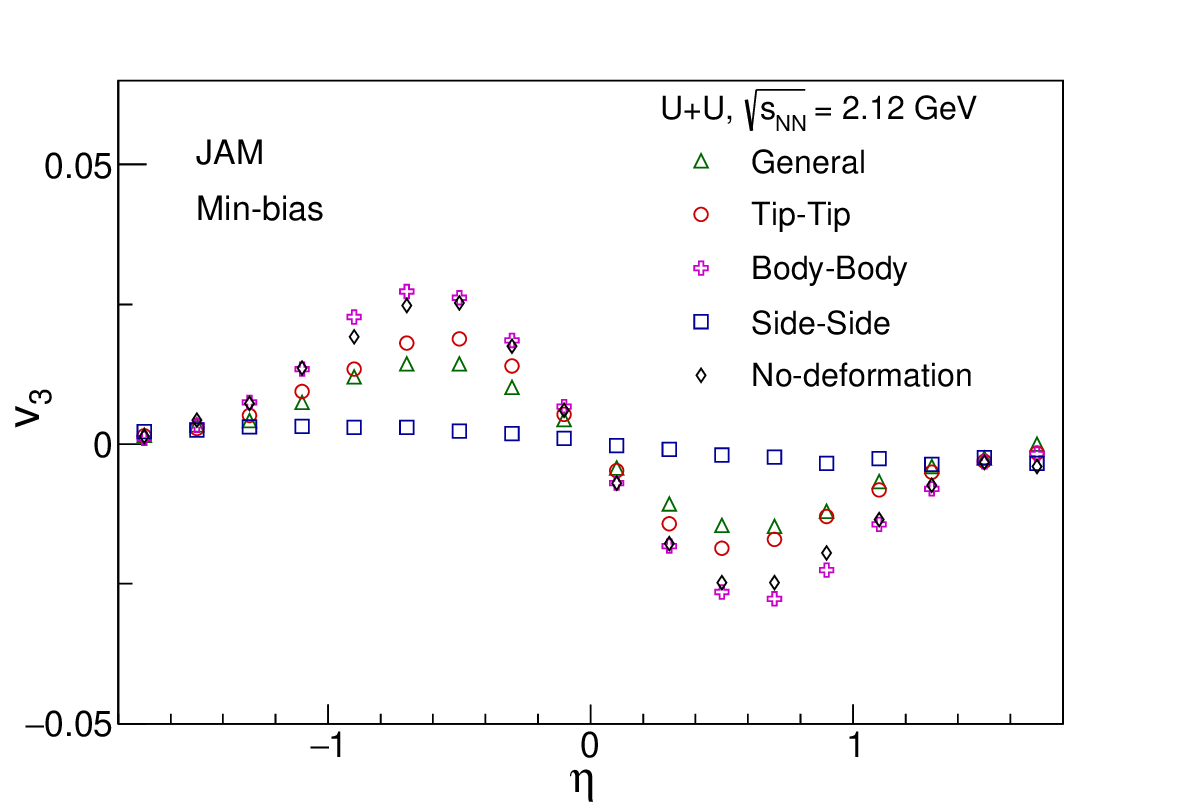}
\caption{ame as Fig.~\ref{fig5_v1etapt} but for triangular flow ($v_{3}$).}
\label{fig7_v3etapt}
\end{center}
\end{figure}

The HADES experiment~\cite{HADES:2020lob}, followed by the STAR experiment~\cite{STAR:2023qnl}, has identified a distinct and substantial negative value for the magnitude of $v_{3}$ of protons relative to the reaction plane. Such $v_{3}$ is found to be anti-correlated with $v_{1}$, and its origin is expected to be from the triangular shape of the participant nucleons as a combined effect from stopping and nuclear geometry. JAM simulations suggested that a mean-field potential is required to describe STAR $v_{3}$ results~\cite{STAR:2023qnl}. In Figure~\ref{fig7_v3etapt}, the top and bottom panels illustrate the $v_{3}$ correlated with the reaction plane, as a function of $p_{\mathrm T}$ and $\eta$. Our observations reveal a substantial $v_{3}$ when the system is in a body-body configuration, displaying the highest magnitude, while it is notably smaller in the side-side configuration.

\begin{figure}
\begin{center}
\includegraphics[scale=0.4]{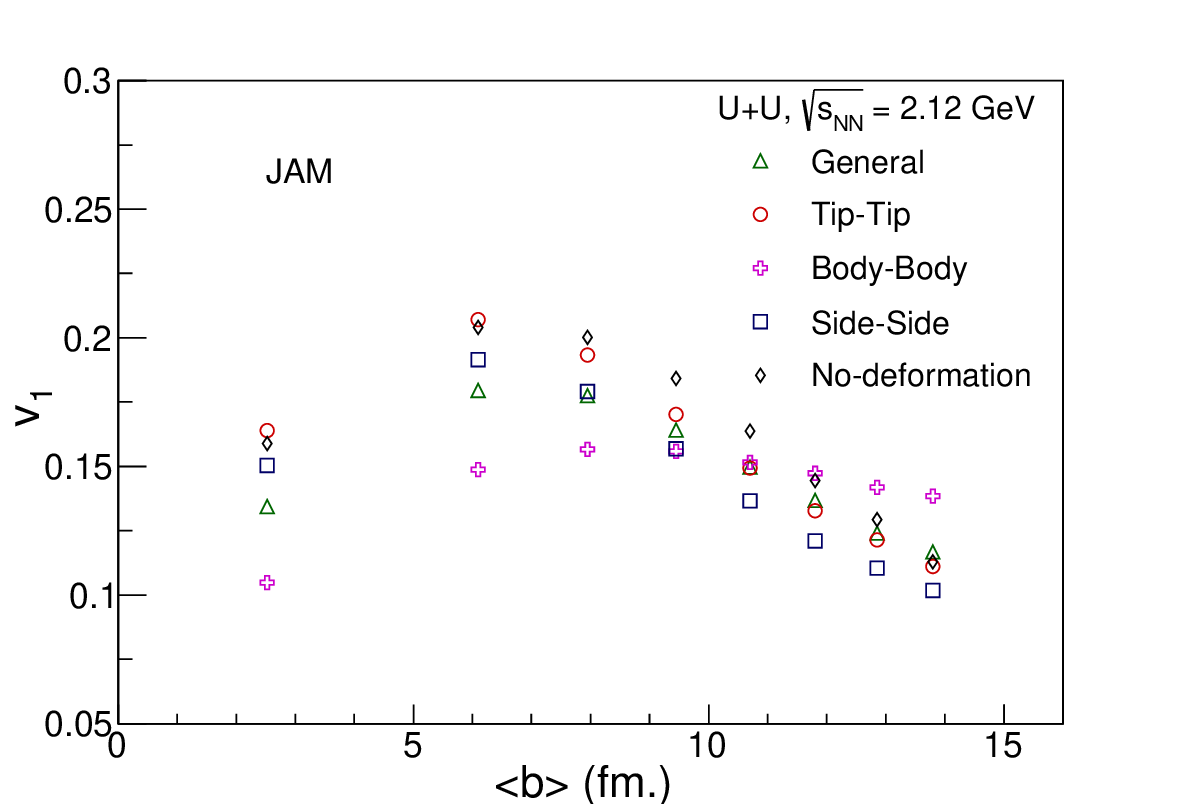}
\includegraphics[scale=0.4]{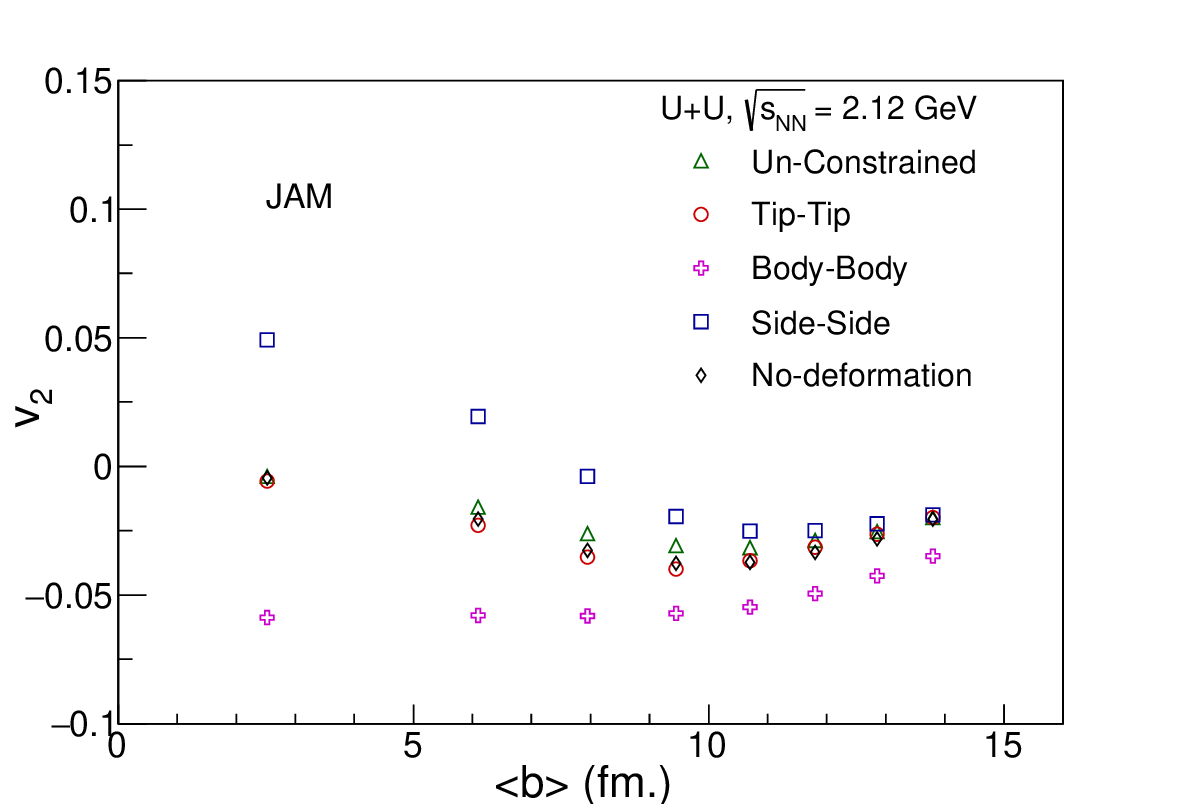}
\includegraphics[scale=0.4]{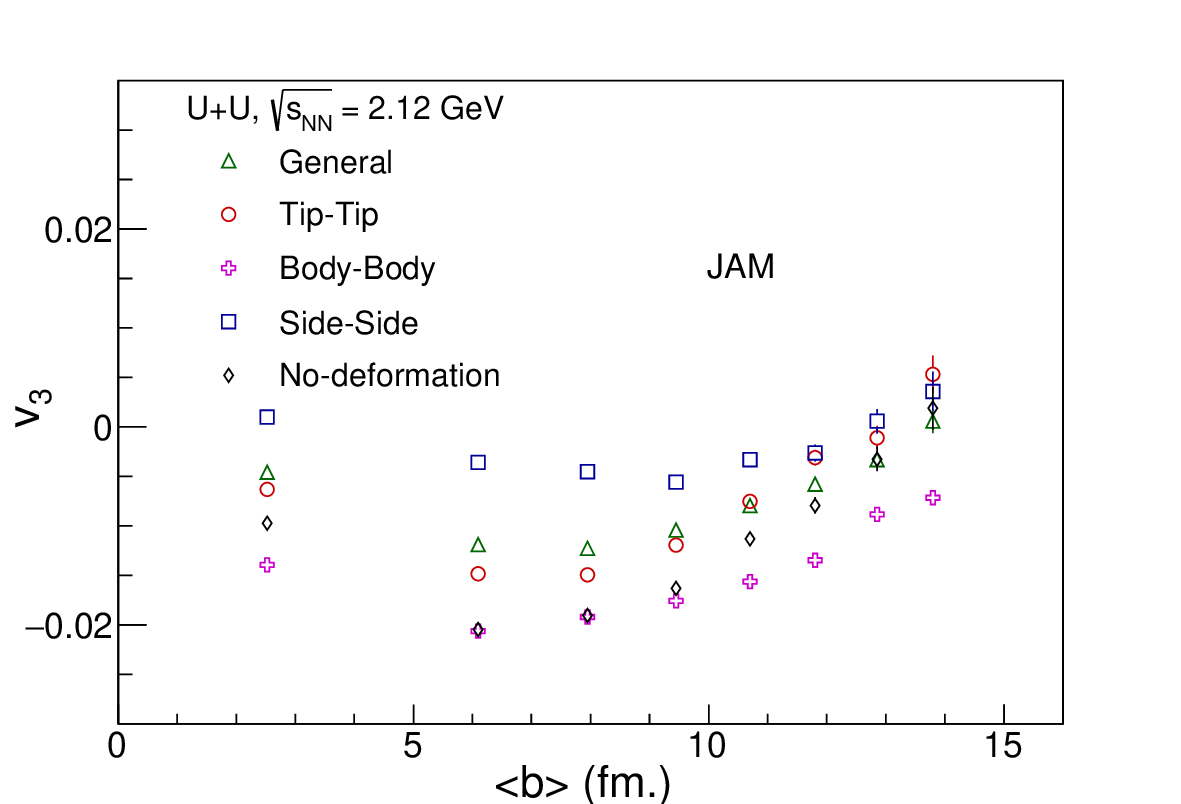}
\caption{ $v_{1,2,3}$ as a function of impact parameter in U+U collisions at $\sqrt{s_{\mathrm NN}}$ = 2.12 GeV for different orientations of colliding nuclei using JAM model.}
\label{fig8_v1cent}
\end{center}
\end{figure} 
The top, middle and bottom panels in Fig.~\ref{fig8_v1cent} presents the impact parameter dependence of ($p_{T}$, y) integrated $v_{1}$, $v_{2}$, and $v_{3}$ respectively (the sign of rapidity is weighted for $v_1$ and $v_3$ in the integration). A strong dependence of $v_{n}$ on $\langle b \rangle$ is observed for all U+U orientations. Notably, a distinct pattern emerges for $v_{1}$ in the body-body configuration, with the lowest values observed in the most central collisions and the highest values in the most peripheral collisions. Conversely, the side-side configuration exhibits the smallest values in the peripheral bins. Across all impact parameters, the magnitude of $v_{2}$ is consistently higher in the side-side configuration compared to the body-body configuration, which consistently displays the lowest magnitude. In the case of the most central collisions, the magnitude of $v_{3}$ is maximum in the side-side configuration, while it remains lowest in the body-body configuration. In peripheral collisions, the magnitude of $v_{3}$ in the body-body configuration continues to be lower, while no clear distinction can be made for the other configurations.

\begin{figure}
\begin{center}
\includegraphics[scale=0.4]{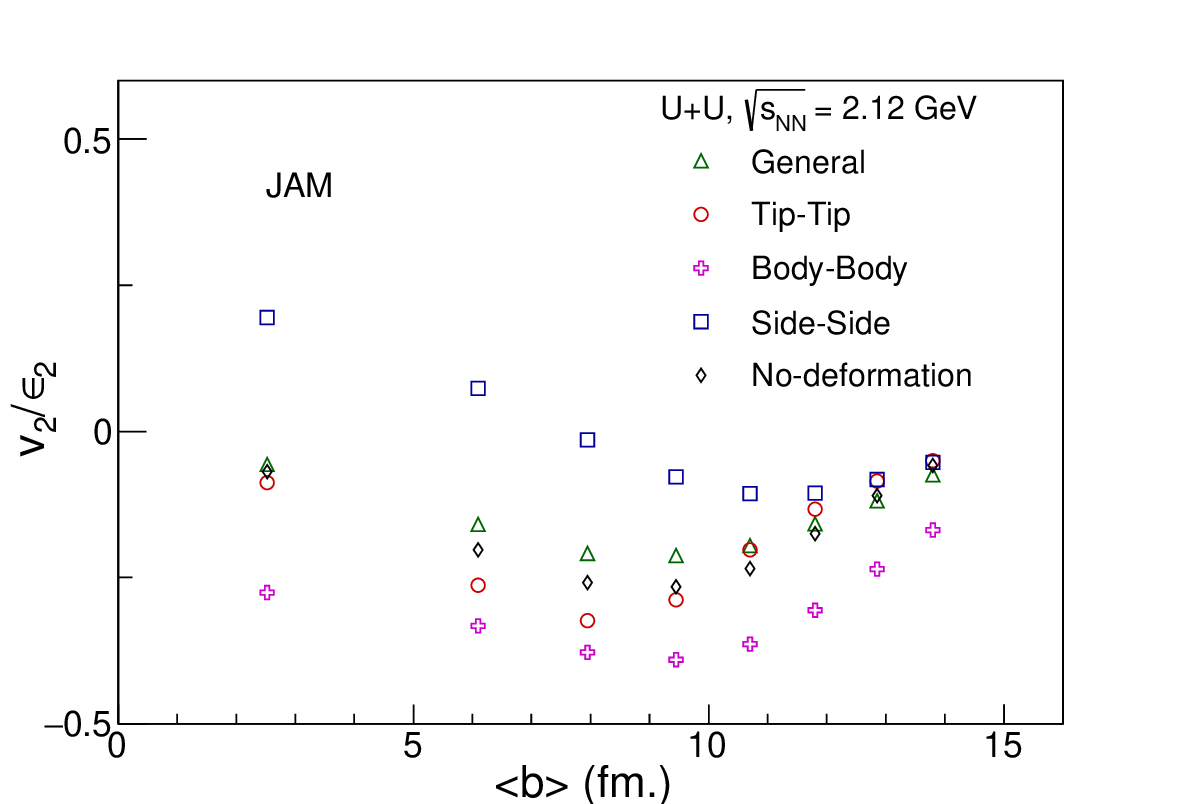}
\includegraphics[scale=0.4]{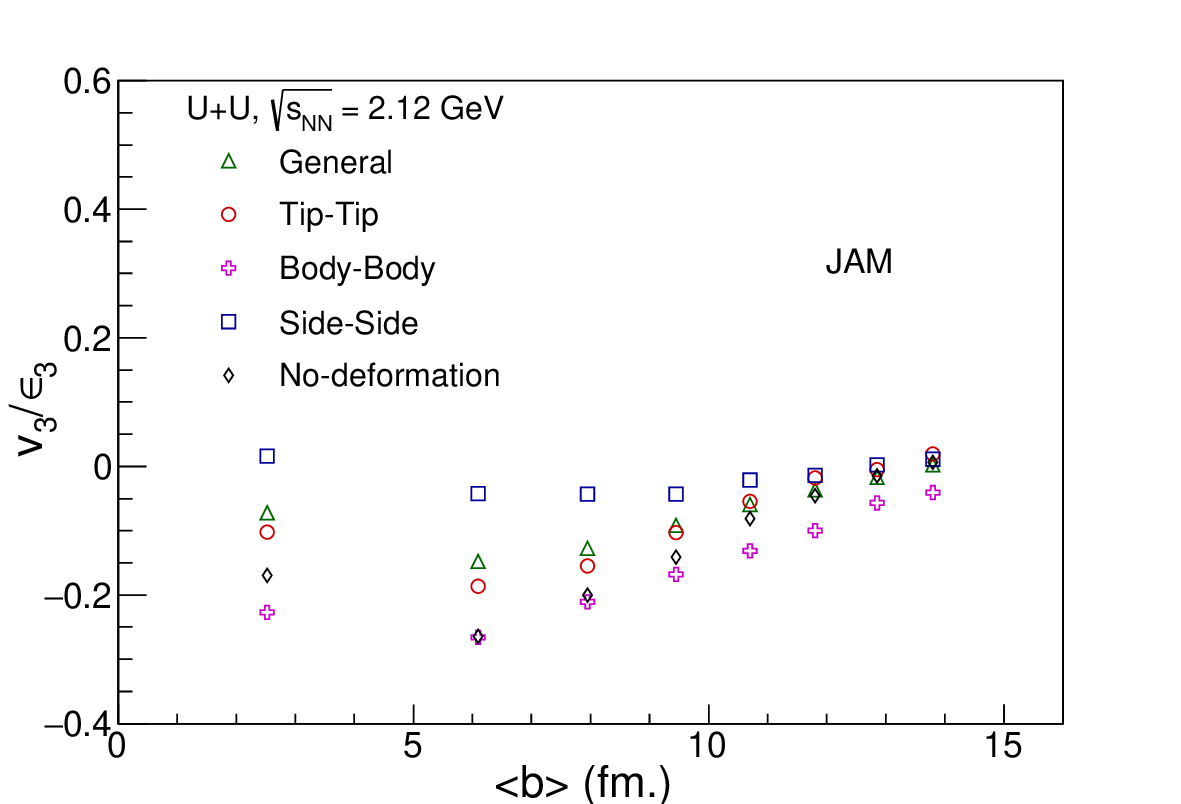}
\caption{ $v_{2}/\epsilon_{2}$  and $v_{3}/\epsilon_{3}$ as a function of impact parameter in U+U collisions at $\sqrt{s_{\mathrm NN}}$ = 2.12 GeV for different orientations of colliding nuclei using JAM model.}
\label{fig9_vnen}
\end{center}
\end{figure}

To disentangle the influence of the initial spatial configuration, we examine the ratio $v_{n}/\epsilon_{n}$. The figures presented in Fig.~\ref{fig9_vnen} consist of two panels, depicting the behavior of $v_{2}/\epsilon_{2}$ and $v_{3}/\epsilon_{3}$ in relation to the impact parameter. While a complex pattern emerges in the case of $v_{2}/\epsilon_{2}$, the behavior of $v_{3}/\epsilon_{3}$ scales perfectly with the expectation from the initial geometric pattern.

\section{Summary and Conclusion}
In summary, we have studied charged particle multiplicity, average $\langle p_{\text{T}}\rangle$, and flow harmonics ($v_{1,2,3}$) for different orientations of deformed Uranium on Uranium collisions at Lanzhou-CSR energy $\sqrt{s_{\mathrm NN}} = 2.12$ GeV employing the JAM transport model. Among the various orientation setups at this energy level, the tip-tip scenario stood out with the highest average charged particle multiplicity. Notably, as we explored different configurations, the second and third-order eccentricities, $\epsilon_{2,3}$, revealed complex patterns in their variations with impact parameters. On the flow harmonics, specifically in the tip-tip configuration, $v_{1}$ exhibited the most significant positive magnitude, while in the body-body configuration, it showed the least pronounced magnitude. When it comes to $v_{2}$, all configurations except the side-side displayed a negative sign consistent with the expectation from dominant in-plane flow compared to out-of-plane. On the other hand, side-side configuration uniquely displayed a positive sign for $v_{2}$. The eccentricity scaled $v_{2}$ shows a rich pattern as a function of collision centrality. When looking at individual $v_{3}$ and $v_{3}/\epsilon_{3}$, there is significant dependency observed across the different orientation scenarios. Our study will provide a baseline understanding of different configurations of U+U collisions at the upcoming CEE experiment. In a prior study~\cite{Giacalone:2021udy}, a linear correlation between $\langle v_{2}^{2} \rangle$ ($\langle \epsilon_{2}^{2} \rangle$) and quadrupole deformation ($\beta_{2}^{2}$) is established in Au+Au collisions at 200 GeV. It facilitates the extraction of $\beta_{2}$ in heavy-ion collisions. In our forthcoming work, we intend to explore these associations by systematically varying the $\beta_{2}$ values within JAM simulations at intermediate energy ranges. Additionally, we plan to employ event-shape engineering techniques to distinguish particular U+U collision orientations from unbiased scenarios.

\section{Acknowledgments}
Financial assistance from Chinese Academy of Sciences is gratefully acknowledged. We would like to thank our colleagues at the Institute of Modern Physics for many insightful discussions.

\bibliographystyle{apsrev4-1}
\bibliography{reference}
\end{document}